# LITHIUM IN STELLAR ATMOSPHERES: OBSERVATIONS AND THEORY

## L. S. Lyubimkov

Of all the light elements, lithium is the most sensitive indicator of stellar evolution. This review discusses current data on the abundance of lithium in the atmospheres of A-, F-, G-, and K-stars of different types, as well as the consistency of these data with theoretical predictions. The variety of observed Li abun-dances is illustrated by the following objects in different stages of evolution: (1) Old stars in the galactic halo, which have a lithium abundance log ε(Li)=2.2 (the "lithium plateau") that appears to be 0.5 dex lower than the primordial abundance predicted by cosmological models. (2) Young stars in the galactic disk, which have been used to estimate the contemporary initial lithium abundance log ε(Li)=3.2±0.1 for stars in the Main sequence. Possible sources of lithium enrichment in the interstellar medium during evolution of the galaxy are discussed. (3) Evolving FGK dwarfs in the galactic disk, which have lower log ε(Li) for lower effective temperature $T_{eff}$ and mass M. The "lithium dip" near $T_{eff}$~6600 K in the distribution of log ε(Li) with respect to $T_{eff}$ in old clusters is discussed. (4) FGK giants and supergiants, of which most have no lithium at all. This phenomenon is consistent with rotating star model calculations. (5) Lithium rich cold giants with log ε(Li) ≥2.0, which form a small, enigmatic group. Theoretical models with rotation can explain the existence of these stars only in the case of low initial rotation velocities $V_0$<50 km/s. In all other cases it is necessary to assume recent synthesis of lithium (capture *of* a giant planet is an alternative). (6) Magnetic Ap-stars, where lithium is concentrated in spots located at the magnetic poles. There the lithium abundance reaches log ε(Li)=6. Discrepancies between observa-tions and theory are noted for almost all the stars discussed in this review.

Crimean Astrophysical Laboratory, Russian Academy of Sciences, Russia; e-mail: lyub@craocrimea.ru

## 1. Introduction

A number of light chemical elements undergo significant changes in their observed abundances during the first, longest stage of stellar evolution, when hydrogen is burning in a star's core: this is the main sequence (MS) stage. In later stages of evolution, after leaving the MS, the abundances of these elements can undergo further changes. Given their crucial role in understanding the evolution of stars, in astrophysics they are sometimes referred to as key elements,.

These elements include the first eight elements of the periodic table. They are listed in Table 1, together with their atomic numbers and solar abundances log ε(El) from Asplund's review [1]. The abundances are all taken relative to hydrogen, which is the most widespread element in the observable universe. Here log ε(El) is given in a standard logarithmic scale where it is assumed that log ε(H) = 12.00. It should be emphasized that this review discusses the abundances of elements in the atmospheres of stars, which are derived from observations of stellar spectra. The atmospheric abundances of elements can differ significantly from their abundances in the interiors of stars where thermonuclear reactions take place.

The author has published a review "Helium in stellar atmospheres" [2] that examines data on the observed

TABLE 1. List of the Eight Lightest Elements and their
Abundances in the Atmosphere of the Sun [1]

| Element | Atomic number | logε(El) |
|---|---|---|
| H | 1 | 12.00 |
| He | 2 | 10.93[*] |
| Li | 3 | 1.05 |
| Be | 4 | 1.38 |
| B | 5 | 2.70 |
| C | 6 | 8.43 |
| N | 7 | 7.83 |
| O | 8 | 8.69 |

* No helium lines are observed in the photospheric spectrum of the sun. The helium abundance given here corresponds to the average abundance for nearby young B-stars [2].

helium/hydrogen ratio (the first two elements in Table 1). The present review is devoted to lithium, the third element in Table 1 and the most sensitive of the elements listed there to stellar evolution. This is explained by the fact that Li is destroyed in (p, ) reactions, for which a temperature of $T \sim 2.5 \times 10^6$ K is sufficient. This means that even slight mixing in the surface layers of a star (perhaps convection in the case of cold stars) can lead to a significant reduction in the Li abundance in the star's atmosphere as it evolves. For example, the amount of Li in the sun's atmosphere has fallen by a factor of 140 during its lifetime (see below).

According to current cosmological models, the first lithium in the universe was formed in the Big Bang. The Big Bang produced the five lightest elements: H, He, Li, Be, and B. In terms of mass and number of atoms, the overwhelming majority were H and He. The primordial helium/hydrogen ratio (in terms of the numbers of atoms) was He/H = 0.082±0.002 [2]. The primordial amounts of Li, Be, and B were much smaller: $^7\text{Li/H} \sim 10^{-10}$, $^9\text{Be/H} \sim 10^{-18}$, and $^{11}\text{B/H} \sim 10^{-18}$ [3]. Here the most abundant isotopes of Li, Be, and B are listed. We note that the current abundances of these three elements are still fairly low (see Table 1). As noted above, however, the amount of lithium in the sun's atmosphere has fallen by a factor of 140 during its lifetime, so the Li abundance given in Table 1 can scarcely be regarded as a standard.

After the formation of the first stars (roughly 400 million years after the Big Bang), the synthesis of heavier elements began in the stars, with simultaneous changes in the amounts of Li, Be, and B. The rather variegated picture of the abundances of Li, Be, and B currently observed is a consequence of two factors: (1) The chemical evolution of the galaxy as a whole, i.e., the formation of new generations of stars in it and the ejection into the interstellar medium of matter from these stars that has been reprocessed in thermonuclear reactions. (2) Internal processes in each star as it evolves.

Lithium is easily ionized (its ionization potential is 5.39 eV), so Li I lines can be seen only in relatively cold stars with effective temperatures $T_{eff}$ < 8500 K, i.e., stars ranging from late A to M. In hotter stars, all the lithium in their atmosphere is fully ionized (i.e., is in the Li II state), so Li I lines cannot be observed. The strong Li I resonance line at a wavelength of 6707.8 Å is most often observed in the spectra of cold stars. Most data on the abundance of lithium in stars is obtained by analyzing this line. In a few, comparatively rare cases, the fainter, subordinate Li I 6103.6 Å line can also be seen.

It should be noted that calculations of Li I lines, including the resonance 6707.8 Å line, usually require rejection of the assumption of LTE (local thermodynamic equilibrium). This is important for an exact determination of the amount of Li and for further comparison with theoretical predictions. It has been said that non-LTE corrections to the lithium abundance log ε(Li) depend on a number of parameters of a star, including its effective temperature $T_{eff}$, acceleration of gravity log$g$, metallicity index [Fe/H], microturbulence velocity $V_t$, and loge(Li) itself [4]. The results of non-LTE analyses of lithium lines (especially where deviations from LTE are substantial) are given preference in discussions of data on lithium in this review.

This review of modern data on the abundance of lithium in stellar atmospheres is concerned with stars in classes F, G, and K, with occasional mention of late A-stars. It should be emphasized that when we speak of the lithium abundance in this review, we usually refer to the abundance of $^7$Li, the most widespread isotope of lithium in the observable universe. In a few, rather rare cases, the less abundant isotope $^6$Li is mentioned.

This review begins with the oldest stars belonging to the galactic halo (population II). These objects can provide information on the primordial abundance of lithium after the Big Bang. Next, young, incompletely evolved stars in the thin disk (population I) are discussed: they can be used to estimate the initial lithium abundance log ε(Li) for stars at the start of their evolution in the MS. Later, the observed evolutionary changes in log ε(Li) in the atmospheres of dwarfs in the disk are examined and then, in the atmospheres of giants and supergiants. The comparatively sparse group of lithium-rich giants and supergiants is discussed separately. Finally, magnetic Ap-stars, in which the lithium lines have interesting behavior, are discussed.

## 2. Lithium in old stars in the galactic halo

Since primordial lithium was produced in the Big Bang according to cosmological model predictions, a survey of current data on the abundance of lithium should logically begin with the following question: what is known about the amount of this element in the atmospheres of the oldest stars? Do the predictions of the Big Bang theory agree with these data.

In order to answer these questions it is necessary to examine stars in the galactic halo (population II). It is known that the evolution of massive stars in the halo ended long ago with the formation of black holes, neutron stars, and white dwarfs. At present, it is only possible to observe mostly old, low-mass stars in this region, predominantly dwarfs with low metal contents (recall that elements heavier than hydrogen and helium are typically referred to as metals in astrophysics).

The metallicity of stars is often characterized by the quantity $[\text{Fe/H}] = \log\varepsilon(\text{Fe}) - \log\varepsilon_\odot(\text{Fe})$, where $\log\varepsilon(\text{Fe})$ and $\log\varepsilon_\odot(\text{Fe})$ are the abundances of iron in the star and in the sun, respectively. The abundance of Fe in the sun is taken to be $\log\varepsilon_\odot(\text{Fe}) = 7.50$ [1]. According to modern ideas, the first stars contained no metals, i.e., they had metallicity [Fe/H]=0 (an extremely small contribution from Li, Be, and B can be neglected). Thus, when [Fe/H] is smaller, we are observing older stars. For stars in population II, a metallicity [Fe/H]<-1 is typical; note that today, old stars with [Fe/H]~ −5 and −6 have been observed in the halo (see below).

The first estimates of the lithium abundance for stars with [Fe/H] from −1.5 to −3.5 led to an unexpected conclusion: these stars had essentially the same lithium abundance $\log\varepsilon(\text{Li}) \approx 2.1$. These values of logε(Li) had no dependence on the metallicity index [Fe/H]. This phenomenon (i.e., the constancy of logε(Li) in this range of [Fe/H]) came to be known as the "lithium plateau" or the "Spite plateau," from the names of the two astronomers who first discovered this interesting effect more than 30 years ago. Initially they found a value of $\log\varepsilon(\text{Li}) = 2.05 \pm 0.15$ [5] for the lithium plateau; the current value is log (Li) = 2.2 [6]. The latter value is considered to be the initial abundance of lithium in the oldest stars in the galaxy.

It is interesting to note that this value has recently been confirmed in studies of cold stars in another stellar system, the Sagittarius dwarf galaxy. An initial lithium abundance of log (Li) = 2.3 was found [7] for stars in the globular cluster M54 in that galaxy, in good agreement with the initial value given above for old stars in our galaxy.

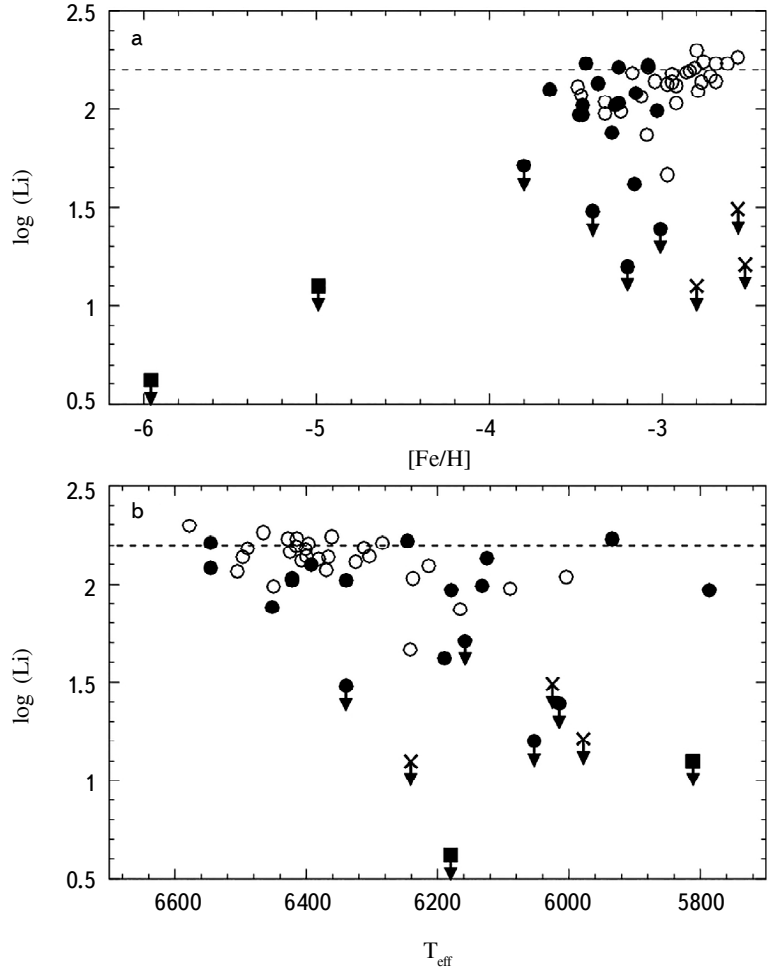

Fig. 1. The lithium abundance in stars in the galactic halo as functions of (a) the metallicity index [Fe/H] and (b) the effective temperature $T_{eff}$ (from the data of Bonifacio, et al. [12]). The symbols with arrows pointing downward correspond to the upper limit of log (Li). The dashed line corresponds to log (Li) = 2.2 (the "lithium plateau").

The discovery of the lithium plateau effectively led to the assumption that this is the primordial lithium created as a result of the Big Bang, but this assumption came upon some serious difficulties. First, it turns out the standard model of the Big Bang (SMBB) predicts a higher abundance of lithium, log (Li)=2.7 [8], than the lithium plateau. Thus, there is a discrepancy of about 0.5 dex between the lithium plateau and the predictions of the SMBB.

In discussing the predictions of the SMBB, one should mention the primordial isotope ratio $^6$Li/$^7$Li for lithium. Recent calculations [9] based on the SMBB yielded primordial isotopic abundances of $\log\varepsilon(^7\text{Li}) = 2.72$ and $\log\varepsilon(^6\text{Li}) = -1.91$; this gives $^6\text{Li}/^7\text{Li} = 2.3 \cdot 10^{-5}$ (for the sun, $^6\text{Li}/^7\text{Li} = 0.08$ [1]). A ratio $^6$Li/$^7$Li as small as this cannot be detected using the observed profile of the Li I 6707.8 Å line. Lind, et al. [10], have made a detailed non-LTE study of the Li I 6707.8 Å profile for four stars in the halo and found no signs of the isotope $^6$Li, thereby confirming the calculations of Ref. 9. As they pointed out [10], in some earlier papers it was reported that $^6$Li had been detected in some old stars, but those results are now in doubt.

The above discrepancy of ~0.5 dex in the values of log ($^7$Li) for the lithium plateau and predicted by the SMBB could be explained in principle by two causes: imperfections of the SMBB and/or systematic errors in determining the abundances log (Li). As for the accuracy of the SMBB, it has been noted [6,8] that this standard model gives primordial abundances of deuterium and the helium isotopes $^3$He and $^4$He in full agreement with the observed values. The outstanding agreement between the observed and cosmological values of the primordial abundance of He has also been noted in Ref. 2. It turns out that a disagreement between the SMBB and observations exists only for lithium. As for possible errors in the observed values of log (Li), however, it would be hard to explain why, with these errors, most of the stars in the halo with their different temperatures $T_{eff}$ yield the same value $\log\varepsilon(\text{Li}) \approx 2.2$, rather than the expected spread in the values of log (Li).

It is also possible that there are no errors, i.e., there actually is a difference of –0.5 dex between the primordial abundance of Li formed by the Big Bang and the actual log (Li) = 2.2 in most of the old dwarfs in the halo. In other words, some processes in the atmospheres of these stars could lower the primordial Li abundance. Korn, et al. [11]. have examined the diffusion of Li atoms in the upper layers of stars as such a process.. Their calculations show that this, together with turbulent mixing, can explain at least half (–0.25 dex) of the discrepancy between the SMBB and the lithium plateau.

Another enigma is that the observed values of log (Li) for some old stars in the halo lie well below the lithium plateau. That is, the distribution of log (Li) with respect to the metallicity index [Fe/H] for these stars is not as simple as initially predicted. This can be seen clearly in Fig. 1, which is based on data from Ref. 12 and references cited there (see Figs. 11 and 12 in Ref. 12). We note that the Li I 6707.8 Å line has been analyzed [12] on the basis of non-LTE calculations. As Fig. 1a shows, over a wide range of [Fe/H] from –2.5 to –6, the series of dwarfs whose spectra do not include the Li I line yield an upper limit of $\log\varepsilon(\text{Li}) < 1.5$. Two stars with extremely low metallicity are of special interest: .SDSS J102915+172927 with [Fe/H]=–5.0 [13] and HE 1327-2326 with [Fe/H]=–6.0 [14] (the two black squares in Fig. 1). The upper limits on the Li abundance for them are $\log\varepsilon(\text{Li}) < 1.1$ and $\log\varepsilon(\text{Li}) < 0.62$, respectively. It is interesting that the stars with lithium abundances substantially below the lithium plateau do not manifest any systematic differences from the other stars in terms of the effective temperatures $T_{eff}$ (Fig. 1b). It is also important to note that for many stars in the halo the observed Li abundance remains fairly close to the lithium plateau $\log\varepsilon(\text{Li}) = 2.2$ (dashed line in Figs. 1, a and b).

There is still no generally accepted answer to the question of why some old dwarfs in the halo with low metallicity have a Li abundance substantially below the lithium plateau. It has been proposed, in particular, that these stars are "blue stragglers" (which also do not have lithium in their spectra).

There is also no sure explanation of why there is a discrepancy of -0.5 dex between the observed "primordial" Li abundance (the lithium plateau) and the predictions of the SMBB. Attempts (sometimes quite exotic) have been made to improve the SMBB. It has been pointed out that this discrepancy can be eliminated by assuming that the fundamental constants change with time [15].

Thus, there are still no unique answers to two important questions regarding the Li abundance in old stars in the galactic halo: (1) Why is there a disagreement between the observed primordial lithium abundance and the predictions of the Big Bang theory? (2) Why do a number of dwarfs in the halo have a Li abundance significantly below the lithium plateau? Until reliable answers have been obtained for these questions, the lithium problem for the old stars in the halo can hardly be regarded as solved.

## 3. The initial abundance of lithium in young stars of the thin galactic disk

It is natural to proceed from the oldest stars in the galaxy to the youngest. This refers to stars in the thin galactic disk (population I) where, as opposed to the galactic halo, star formation is ongoing at present. Can the initial abundance of Li in these stars be determined observationally? This is possible if we examine those young stars which have not yet evolved significantly, so that the abundance of Li in their atmospheres has not changed from the initial value.

Studies [16,17] of MS F- and G-dwarfs with a normal (solar) metallicity have shown that the initial lithium abundance for these stars averages $\log\varepsilon(Li) = 3.2 \pm 0.1$. This value for the lithium abundance in cold dwarfs may decrease significantly as they evolve and do so more rapidly when their effective temperatures $T_{eff}$ or masses are lower (see below).

In connection with this problem, there is some interest in the Li abundance in T Tau stars which have not yet reached the zero age main sequence (ZAMS) stage and are, therefore, very young. These are cold dwarfs in classes G, K, and M with low masses ($M \sim 1 M_\odot$ and less). Martin, et al. [18], have determined the Li abundance for 55 stars of this type with masses $M$ from 1.2 to $M_\odot$ and found a sharp maximum in the distribution of loge(Li) at 3.1. They concluded that the initial lithium abundance for T Tau stars is $\log\varepsilon(Li) = 3.11 \pm 0.06$. The actual coincidence of this value with the initial $\log\varepsilon(Li) = 3.2 \pm 0.1$ found above for young MS F- and G-dwarfs is noteworthy. It should be noted that for stars with masses $M < 1.2 M_\odot$ a reduction in the Li abundance (Li depletion) can set in before they reach the MS, i.e., before the start of thermonuclear hydrogen burn in a star's core. The observed amount of Li in T Tau stars confirms this conclusion [18,19].

Young stars in population I, therefore, have an initial lithium abundance of $\log\varepsilon(Li) = 3.2 \pm 0.1$. This is an indirect indication of the same abundance of lithium in the interstellar gas-dust medium from which these stars were formed (at least for stars in the vicinity of the sun with normal metallicity). It is interesting to note the nice agreement between this value and the Li abundance in meteorites, $\log\varepsilon(Li) = 3.26 \pm 0.05$ [1]. The latter value can be regarded as the Li abundance at the start of the formation of the solar system, i.e., 4.5 billion years ago.

It is important that the Li abundance in young stars in population I is an order of magnitude greater than the above mentioned "relict" value log (Li) = 2.2, which is characteristic of old stars in the halo (the lithium plateau). The following question arises: how can this additional lithium show up in the thin disk? The sources of the lithium enrichment of the thin disk (more precisely, of the interstellar medium in the disk) may be stars of the following types [20,21]: supernovae, novae, red giants, and stars in the asymptotic giant branch (AGB). The lithium synthesized in these stars could be ejected into the surrounding interstellar medium by outflows. The isotope $^7$Li could also be synthesized by galactic cosmic rays in spallation reactions with the heavier and much more abundant nuclei of C, N, and O in the interstellar medium.

Modelling the chemical evolution of the galaxy with these processes taken into account [20] has shown that the largest contribution to lithium enrichment (more precisely, in the isotope $^7$Li) up to the observed level in young stars, log (Li)=3.2, was from red giants with masses $1 \leq M/M_\odot \leq 2$. Next in terms of significance are cosmic rays (25%), novae (18%), type II supernovae (9%), and AGB stars (0.5%). These are approximate estimates. In particular, as noted in Ref. 21, the contribution of AGB stars could be lower than in Ref. 20. It has recently been shown [22] that more than 50% of the lithium enrichment in the vicinity of the sun has been from low-mass stars of the following types: red giants, AGB stars, and novae. The generation of cosmic rays in the galaxy and their role in the formation of the isotopes of Li, Be, and B has been studied in detail [22]. In particular, 100% of the isotopes $^6$Li, $^9$Be, and $^{10}$B (and 70% of $^{11}$B) are produced by galactic cosmic rays. The conclusion of Ref. 20 was confirmed: cosmic rays are the source of about 20% of the isotope $^7$Li in the sun's vicinity. (Here the value 18% given in Ref. 22 has been rescaled to the SMBB.)

It should be noted that models of the chemical evolution of our galaxy and other galaxies have been under development for more than 30 years. These models can explain a number of features of the observed abundances of elements in the thin disk and in the halo of the galaxy. The basic principles behind these models are discussed, for example, in the review by Matteucci [23].

## 4. Evolution of the lithium abundance in the atmospheres of dwarfs in the galactic disk

So the initial lithium abundance in young stars near the sun is log (Li) = 3.2. The observed Li abundances for stars in the thin disk may, however, be considerably lower than this value and these variations are related to the evolution of the stars. We begin the discussion of the evolutionary changes in the atmospheric abundance of Li with class F-M dwarfs, which are stars in population I in the MS stage.

The dependence of the atmospheric lithium abundance in F-M dwarfs on $T_{eff}$ (i.e., in fact on mass $M$) and on age is well known: the amount of Li decreases with age and more rapidly when $T_{eff}$ or $M$ are lower. The dependence of log (Li) on $T_{eff}$ can be detected especially well in the stars in a single cluster, since their ages are approximately the same. As an example, Figure 2 [24] shows this dependence for two diffuse clusters with different ages $t$: (a) Hyades ($t \sim 700$ million years) and (b) Pleiades ($t \sim 100$ million years). The trend in the lithium abundance shows up clearly: log (Li) decreases with falling $T_{eff}$, and this trend is steeper in the older Hyades cluster. The latest data on lithium for the Hyades [25] and Pleiades [26] provide reliable confirmation of this trend.

The dependence of log (Li) on $T_{eff}$ gives researchers an independent method for determining the age of close binary systems if both components have the Li I 6707.8 Å line in their spectra. As an example we take the binary system ι Peg (HR 8430), to which a special method described in Ref. 24 has been applied [27], in order to obtain the individual parameters of the components, including $T_{eff}$, $\log g$, and the lithium abundance log (Li). The difference of 0.67 dex in the values of log (Li) for the components yielded an estimated age of $t \approx 200$ million years.

The distribution of log (Li) in the Hyades (Fig. 2a) reveals another interesting feature: a deep "lithium dip" near $T_{eff} \sim 6600$, where the lithium abundance falls by two orders of magnitude. No such dip is observed in the Pleiades; that is, it is characteristic only of rather old clusters. We now examine these two effects (the trend and the dip) in more detail.

The most natural explanation for the reduced lithium abundance ("lithium depletion") in the atmospheres of class F, G, K, and M dwarfs could be mixing owing to convection. Convection is weakly developed in class F dwarfs ($T_{eff} \geq 6000\,K$), but for lower $T_{eff}$, i.e., for G-, K-, and M-stars, the convective shell becomes stronger; thus, when $T_{eff}$ is lower, convective mixing should reduce the amount of Li in the atmosphere more rapidly. This could explain the observed trend in log (Li) with $T_{eff}$ for stars in a single cluster (Fig. 2). Calculations show, however, that convection in the MS stage is not enough to explain the observed reduction in the lithium abundance.

Let us consider the sun from this standpoint. Current estimates of the Li abundance give $\log\varepsilon(Li) = 1.05 \pm 0.10$

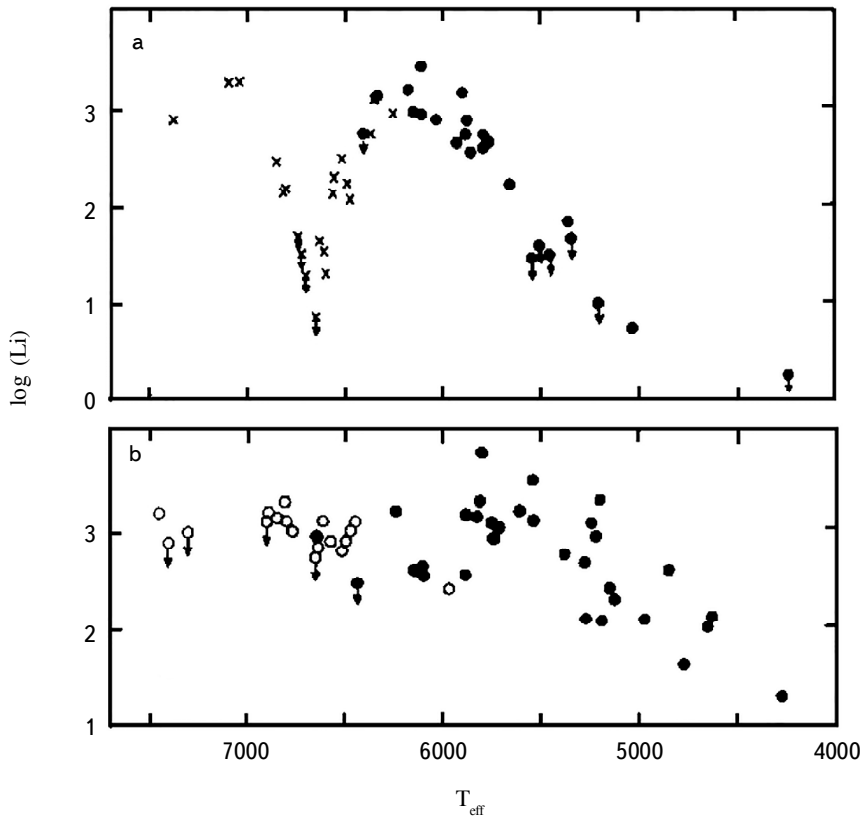

Fig. 2. Lithium abundance as a function of $T_{eff}$ for stars in two clusters with different ages: (a) Hyades ($t\sim 700$ million years) and (b) Pleiades ($t\sim 100$ million years) [24].

[1] and 1.07±0.02[28]. Thus, during the sun's lifetime of $t \sim 4.5 \times 10^9$ years, the Li abundance in its atmosphere has fallen by 2.1 dex relative to the primordial $\log\varepsilon(\text{Li}) = 3.2$. Calculations show that convection cannot explain the observed reduction in log (Li) in the sun's atmosphere, since even in the convective zone of the sun, the temperature does not reach the $T \sim 2.5 \times 10^6$ K required for the initiation of lithium burnup.

Attempts have been made to improve the convection model for the sun in order to extend convective mixing to deeper layers. In addition, additional mixing mechanisms that would further deplete Li in the atmosphere have been proposed for the sun and other dwarfs, such as gravitational diffusion, rotation, mass loss, and internal gravitational waves [29]. Lithium depletion may also be associated with the formation of planetary systems; it turns out that stars with planets have a tendency toward greater lithium deficits [30].

What kind of changes in logε(Li) can we expect for the sun in the future? This question is answered to some extent by a non-LTE comparison [28] of the abundances of Li for the sun ($t \sim 4.5 \times 10^9$ years) and its older twin, the star HIP 102152 ($t \sim 8.2 \times 10^9$ years). This yielded $\log\varepsilon(\text{Li}) = 1.07 \pm 0.02$ for the sun and $\log\varepsilon(\text{Li}) = 0.48 \pm 0.07$ for HIP 102152. The difference in the lithium abundances was about 0.6 dex. Hence, the Li abundance in the sun should decrease by 0.6 dex over the next $3.7 \times 10^9$ years.

Another important phenomenon in Fig. 2 requires an explanation: the "lithium dip" observed in F-dwarfs in the Hyades cluster and in other relatively old clusters with ages $t > 200$ million years. It should be pointed out again that the dependence on $T_{\text{eff}}$ in Fig. 2 actually reflects a dependence on the mass $M$. The location of the "lithium dip" in different clusters depends on their metallicity index [Fe/H]: the dip shifts toward lower $T_{\text{eff}}$ and $M$ when [Fe/H] is reduced. In particular, the center of the dip for the Hyades cluster with a normal metallicity corresponds to $M = 1.4 M_\odot$, while the center of the dip corresponds to $M < 1.06 M_\odot$ for the metal-deficient ([Fe/H] = − 0.54) cluster NGC 2243 [31].

It has been shown [32] that the "lithium dip" may originate in internal gravitational waves near the base of the convective zone. These waves interact with the star's rotation and influence large-scale mixing (meridional circulation and turbulence) which leads to transport of lithium atoms into deep layers and their burnup there. It is important that the distribution of the rotation velocities $V_{rot}$ of the dwarfs with respect to $T_{\text{eff}}$ undergoes a sharp drop in the region of the dip from roughly 150 to 10 km/s (see Fig. 1 in Ref. 32). Hydrodynamic models in which all these effects are taken into account can explain [32] the "lithium dip".

## 5. Lithium in the atmospheres of cold giants and supergiants

On completing the MS stage, dwarfs and subgiants (stars in luminosity classes V and IV) rapidly move into the next evolutionary stage – the giant and supergiant stage (luminosity classes III, II, and I). The giants and supergiants in classes F, G, and K, in whose spectra lithium can be seen, include stars with different masses $M$. The relatively massive stars of this type with $M = 3 - 20 M_\odot$ are the successors to the MS B-stars, while the less massive stars with $M \leq 2 M_\odot$ are the successors of dwarfs in class A and later classes.

The theory predicts that the evolutionary phase of FGK-giants and supergiants is accompanied by deep convective mixing, which causes significant changes in the observed abundances of several light elements. The observed abundance of nitrogen increases and that of carbon decreases during the convective mixing (CM) phase, while the lithium abundance in the atmosphere falls to undetectable levels. Note that the changes in the Li abundance during the CM phase begin earlier than the changes in the abundances of C and N. For a better understanding of the data on the lithium abundance, we now examine briefly the nitrogen and carbon anomalies in these stars and their current interpretation.

**5.1. Nitrogen and carbon anomalies.** The anticorrelation between the abundances of nitrogen and carbon in AFGK-giants and supergiants has been known for more than 30 years. On the average, when the excess of N is larger, the deficit of C is greater (for example, see Fig. 6.4 of Ref. 24). Comparisons with theory were limited during those years by the fact that only models of stars without rotation could be calculated. Later it turned out that rapid rotation in sufficiently massive stars can greatly influence their evolution, including the elemental abundances in the atmosphere [33]. Only modern models with rotation have been able to explain the observed anomalies in N and C in AFG-supergiants and giants both qualitatively and quantitatively, including the anticorrelation between N and C [34,35].

These model calculations [36,37] have been compared [35]. This comparison yielded an important conclusion: the observed anticorrelation between N and C in these stars primarily reflects the dependence of the N and C anomalies on the initial rotation velocity $V_0$; when $V_0$ is higher, these anomalies are greater (see Fig. 12 of Ref. 35). This implies that the initial rotation velocity $V_0$ is just as important as the mass $M$ in the evolution of these stars.

It is important that in a rotating star, changes in the atmospheric abundances of N, C, and other light elements (in particular, Li) can even begin in the first stage of evolution, the MS stage. In fact, if $V_0$ is high enough, the rotation induces mixing in this early stage and it causes transfer of fusion reaction products from the depths of the star to its surface. More than 30 years ago, the author detected an increased abundance of N with age in early MS B-stars, the predecessors of the AFG-supergiants and giants [38].

The following conclusions of Ref. 35 are of interest to us. The observed deficit of C (up to -0.7 dex), excess of N (up to +0.9 dex), and elevated N/C ratio (up to +1.6 dex) indicate that mixed matter is present at the surface of most of the AFG-supergiants and giants. Some of these stars have undergone mixing in the MS stage owing to rotation with initial velocities $V_0 \approx 200-300$ km/s. Others have also undergone later convective mixing (CM) in the AFG-supergiant stage: for these stars $V_0 \approx 0-150$ km/s. Two supergiants (HR 825 and HR 7876) with $V_0 \sim 0$ km/s have already entered the MS, but have not yet reached the CM phase; thus, they have essentially retained their initial C and N abundances. One supergiant (HR 1865) has definitely passed the CM phase and a comparison with the models gives an estimate of $V_0 \approx 300$ km/s for it.

Thus, a comparison of the observed abundances of C and N in AFG-supergiants and giants with rotating star models yields a range of initial rotation velocities $V_0$ from 0 to 300 km/s. The evolutionary status of these stars allows two possibilities: (1) they have already left the MS but not passed through the CM phase to the late giant stage (this is the "post-MS" variant); or (2) they have already passed, even if partially, the CM stage (this is the "post CM" variant). Based solely on the observed abundances of C and N, for many AFG-supergiants and giants it has not been possible to make a unique choice between these two variants. Knowledge of the lithium abundance can help in making this choice.

**5.2. Lithium abundance.** It turns out that the overwhelming majority of AFG-giants and supergiants do not have lithium in their spectra. For example, of 378 G- and K-giants [39], the Li I 6707.8 Å line was detectable in the spectra of only about 10%, while of 417 red giants and supergiants in the galactic bulge [40], this line was observed in only about 3%. We may, therefore, conclude that there is little lithium in the atmospheres of these stars or it has completely burnt up during their evolution.

An explanation of this phenomenon was given almost 40 years ago in the first papers on massive determination of the Li abundance in stars of this type [41,42]. It was proposed that the reason for the observed Li deficit is the above mentioned deep mixing in this stage of evolution (CM phase), because of which lithium falls into deep hot layers and is burnt up there. As noted above, at that time only models without rotation existed. A detailed comparison of the observed Li abundances with modern theoretical models (with and without rotation) can be found in Ref. 43. It is useful to mention once again that lithium burnup during the CM phase begins earlier than the changes in the abundances of nitrogen and carbon.

This non-LTE analysis [43] of the abundances of Li for 55 F- and G-supergiants and giants (stars in luminosity classes I, II, and III) and a comparison with the predictions of the modern theory showed that the atmospheric abundance of Li depends strongly on a star's mass $M$ and initial rotation velocity $V_0$. The masses of the stars found in Ref. 43 using their evolutionary tracks varied from 2 to $15 M_\odot$. It was concluded that the stars can be broken up into two groups with masses $M < 6 M_\odot$ and $M > 6 M_\odot$ on the basis of the observed Li abundances, as well as on the basis of a comparison with theory.

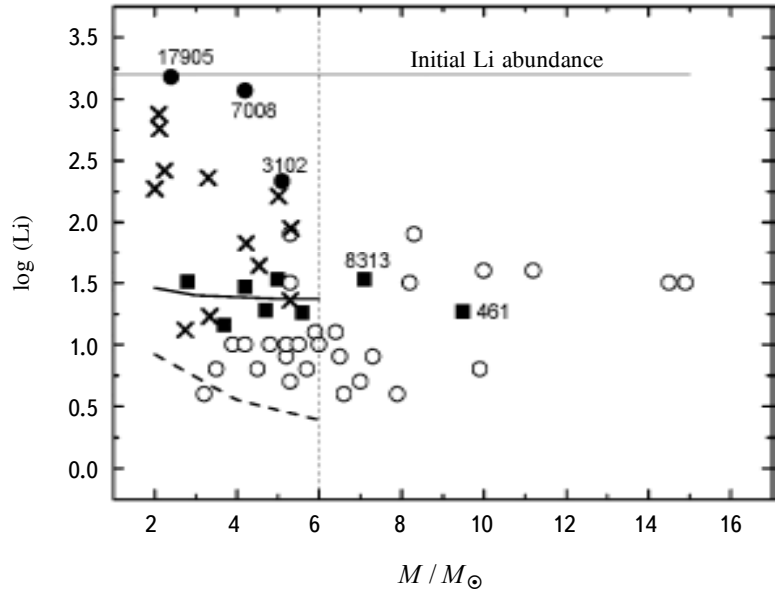

Fig. 3. Lithium abundance in the atmospheres of AFG supergiants and giants as a function of their masses [43]. The open circles correspond to an upper bound on log (Li).

Figure 3 [43] shows the observed lithium abundances $\log\varepsilon(Li)$ of these giants and supergiants, as well as some results of the modern stellar models [44,45]. The thick curve in the region $M = 2 - 6 M_\odot$ corresponds to the theoretical predictions for the "post-CM" phase without rotation ($V_0 = 0$). The dashed curve below it corresponds to the models with $V_0 \approx 50$ km/s.

The above-mentioned partition of stars with masses $M$ above and below $6 M_\odot$ is quite clear in Fig. 3. On one hand, for stars with $M < 6 M_\odot$ there is a large scatter in the observed values of log (Li), which range from an initial abundance $\log\varepsilon(Li) = 3.2 \pm 0.1$ in HR 17905 and HR 7008 until there are no signs of lithium in the spectra of many of the stars (open circles). On the other hand, for most of the stars with $M > 6 M_\odot$ there is typically no detectable lithium; Lit was found in only two supergiants (HR 461 and HR 8313). It is noteworthy that in this group there are no stars with $\log\varepsilon(Li) \geq 2.0$ (Li-rich stars; see the next section).

An appreciable group of the stars with $M < 6 M_\odot$ in Fig. 3 have lithium abundances $\log\varepsilon(Li) \sim 1.5$ (black squares). This is the value that was predicted in the first papers on modelling the evolution of nonrotating red giants [46,47]. It agrees well with the current estimate, $\log\varepsilon(Li) = 1.4$; this is the average value for the "post-CM" models without rotation ($V_0 = 0$) and masses $M = 2 - 6 M_\odot$ (the continuous broken line in Fig. 3). It is interesting that, even for small initial rotation velocities $V_0 \approx 50$ km/s, the Li abundance decreases by 2.3-2.9 dex toward the end of mixing (dashed curve in Fig. 3) and is difficult to detect. For $V_0 \approx 100$ km/s the decrease in the Li abundance for the models with $M = 2 - 6 M_\odot$ is already 4-5 dex by the end of this stage. With this strong sensitivity of log (Li) to $V_0$, it is understandable why lithium is not observed in the spectra of a majority of the red giants and supergiants

Figure 3 shows that, in the case of stars with masses $M > 6 M_\odot$, the situation is quite different from the case of $M < 6 M_\odot$. The absence of lithium in the atmospheres of most of these giants and supergiants is fully consistent with modern calculations, which indicate a sharp reduction (by several orders of magnitude) in the atmospheric abundance of Li by the end of the MS stage. For the models without rotation ($V_0 = 0$), the Li abundance does not vary before the end of the MS, but mixing takes place immediately after emergence from the MS and leads to a sudden drop in the Li abundance. It has been shown [43] that mixing occurs in the models with $V_0 = 0$ within a narrow range of effective temperatures, specifically for $T_{eff} = 15820$ K in the models with $M = 7 M_\odot$ and for $T_{eff} = 15700$ K in the models with $M = 15 M_\odot$. Thus, the calculations show that for stars with masses $M > 6 M_\odot$ the lithium should essentially be burnt up even before their emergence into the red giant/supergiant phase. Therefore, the modern theory fully explains the absence of lithium in the atmospheres of most of the stars with $M < 6 M_\odot$, as well as with $M > 6 M_\odot$, although the reasons for the absence of Li may be different in these two groups.

In Fig. 3 the supergiants HR 8313 and HR 461 (indicated in the figure) drop decisively out of the overall picture with their quite detectable lithium abundances, $\log\varepsilon(Li) = 1.53$ and 1.27 [43], respectively. The masses of these stars definitely exceed $6 M_\odot$ ($M = 7.1$ and $9.5 M_\odot$), so that according to the calculations for models with $M > 6 M_\odot$, they should not exhibit lithium. Nevertheless, lithium is observed and it might be assumed that it has recently been synthesized in these supergiants. The possible synthesis of Li in FGK-giants and supergiants is discussed in the next section.

Thus, the abundance of lithium is extremely sensitive to the initial rotation velocity $V_0$. For example, if $V_0 \approx 100$ km/s toward the end of the MS stage, then the abundances of C and N in the atmosphere essentially remain unchanged (see Fig. 12 of Ref. 35), while the Li abundance falls by several orders of magnitude and thereby becomes unobservable.

It is clear from the above discussion that modern theoretical models with rotation provide a complete explanation of the absence of lithium in the atmospheres of most FGK-giants and supergiants. These models can explain the presence of detectable lithium for log (Li) < 2. However, stars with $\log\varepsilon(\text{Li}) \geq 2$ are generally difficult to explain in terms of the classical theory. These "Li-rich" stars merit a separate discussion.

## 6. Cold giants and supergiants that are rich in lithium

Stars that are rich in lithium with abundances $\log\varepsilon(\text{Li}) \geq 2$ are of great interest because a majority of them cannot be explained in terms of the standard theory of stellar evolution. These objects form a very small fraction of all FGK-giants and supergiants. For example, of the 145 FGK-bright giants (luminosity class II) in one study [48], only 5 (~3%) were of that type (it should be noted that the LiI 6707.8 Å line was analyzed [48] in an LET approximation). Of 417 red giants in the galactic bulge studied in another paper [40], only 4 (~1%) had log (Li) > 2.0. Of 378 G- and K-giants studied in yet another paper [39], only 3 (~1%) are "Li-rich." The fact that there are few of these stars may indicate that this phase of evolution is short.

In the literature there is a tendency to divide these stars into giants, "Li-rich giants," and "super Li-rich giants." Adopting this division, we define the following ranges of log (Li) for the latter two "rich" groups: log (Li) = 2.0-3.3 in the first and log (Li) = 3.5-4.3 in the second. The fundamental difference in the Li abundances between these two groups is that log (Li) for the "Li-rich" giants does not exceed the initial value log (Li) = 3.2±0.1, while for the "super Li-rich" giants the initial value is considerably exceeded.

**6.1. The "Li-rich" giants.** It should be noted that in Fig. 3 there a quite a few stars with $\log\varepsilon(\text{Li}) \geq 2$. This apparent abundance is a consequence of the fact that giants from a list in Ref. 49 were added to the original list in Ref. 43, but only those 12 stars which had reliably detectable lithium contents (the crosses in Fig. 3). Note that new effective temperatures $T_{eff}$, accelerations of gravity log$g$, and masses $M$ were determined for these twelve stars in Ref. 43. The non-LTE lithium abundance log (Li) was found from the equivalent width of the LiI 6707.8 Å line. These included 7 "Li-rich" stars.

An important fact was established in Ref. 43: all giants and supergiants of the "Li-rich" type have masses $M < 6 M_\odot$. This conclusion also follows from all the available published data.

TABLE 2. Parameters of Ten "Li-Rich" Giants Studied in Ref. 43

| HR | HD | Sp | $d$, pc | $T_{eff}$ | log$g$ | $V$sin$i$, km/s | $M/M_\odot$ | log$\varepsilon$(Li) |
|---|---|---|---|---|---|---|---|---|
| -- | 17905 | F5 III | 157 | 6580 | 3.26 | 53 | 2.4 | 3.18±0.11 |
| 3102 | 65228 | F7 II | 161 | 5690 | 2.17 | 12 | 5.1 | 2.33±0.19 |
| 7008 | 172365 | F8 II | 342 | 6220 | 2.53 | 58 | 4.2 | 3.07±0.12 |
| 1135 | 23230 | F3 II | 170 | 6570 | 2.39 | 49 | 5.0 | 2.21 |
| 1644 | 32655 | F0 III | 236 | 7150 | 2.98 | -- | 3.3 | 2.36 |
| 6707 | 164136 | F5 II | 264 | 6410 | 2.29 | 27 | 5.3 | 1.95 |
| 1298 | 26574 | F0 III | 37 | 7040 | 3.62 | 108 | 2.1 | 2.76 |
| 2936 | 61295 | F1 III | 102 | 6880 | 3.51 | 35 | 2.1 | 2.88 |
| 5913 | 142357 | F4 III | 95 | 6530 | 3.48 | 27 | 2.0 | 2.27 |
| 6604 | 161149 | F1 III | 120 | 6910 | 3.42 | 66 | 2.2 | 2.42 |

As an example, Table 2 gives the parameters of ten Li-rich giants studied in Ref. 43. The lower part of the table shows the above mentioned 7 Li-rich giants from the list of Ref. 9 (the stars from HR 1135 to HR 6604) separately. All the giants in Table 2 are fairly nearby stars, with distances ranging from 37 to 342 pc. The masses $M$ of these giants vary from 2.0 to $M_\odot$, i.e., the above inequality $M < 6M_\odot$ holds. It should be noted that all the Li-rich giants in Table 2 belonged to spectral class F; their effective temperatures $T_{eff}$ range from 5690 to 7150 K. Colder stars are, however, encountered among Li-rich giants. For example, data for more than 30 K-giants with lithium abundances log (Li) = 2.0-3.3 and $T_{eff}$ = 3900-5150K are given in Ref. 50.

Two F-giants, HD 17905 and HR 7008, with log (Li) equal to the initial abundance log (Li) = 3.2±0.1 of young stars (to within the limits of error) show up in Table 2 and Fig. 3. Is it possible that these stars, having reached the cold giant state, retained their initial lithium abundance without change? In the above discussion of the nitrogen and carbon anomalies (section 5.1), two supergiants (HR 825 and HR 7876) were mentioned which had $V_0 \sim 0$ and had already left the MS, but still had not reached the full CM stage; thus, they had essentially retained their initial abundances of N and C. However, in the case of the giants HD 17905 and HR 7008, this scenario obviously does not apply, since these stars have large rotation velocities, ~53 and 58 km/s, respectively (Table 2). It we assume that at the beginning of the MS these two giants had rotation velocities $V_0 \approx 50$ km/s, then calculations [45] show that the lithium abundance in their atmospheres (more precisely, in models with masses 2 and $M_\odot$) toward the end of the MS should fall by 0.6 and 1.1 dex, respectively; but this is not observed. Thus, from the standpoint of the theory, the giants HD 17905 and HR 7008 could not possibly retain their initial lithium abundance unchanged. This suggests that lithium synthesis has recently taken place in these stars.

It can be seen from Table 2 that elevated rotation velocities $V$sin$i$ are generally typical of most Li-rich stars.

It is important that similar values of $V\sin i$ are not at all typical for cold giants with these masses. In fact, calculations [37] show that for stellar models with masses $M = 2-5 M_\odot$ and initial rotation velocities $V_0 \approx 200$ km/s, after reaching the G- and K-giant or supergiant stage (more precisely, at the end of the helium burnup phase), the surface rotation velocities are only 3-5 km/s. This conclusion is in good agreement with observations of normal red giants. For example, the observed velocities $V\sin i$ of most giants with masses of $M = 1.3 - 2.2 M_\odot$ in 11 diffuse clusters were found to be 1.5 km/s [51]. On the other hand, Table 2 includes five Li-rich F-giants with $M \approx 2 M_\odot$ that have much higher values of $V\sin i$ ranging from 27 to 108 km/s (for HR 1298 [52]). It is interesting that, as opposed to the Li-rich stars given in Table 2, for the overwhelming majority of the 33 stars for which lithium was not found in Ref. 43 (for these, only an upper bound on log (Li) could be estimated), the rotation velocities are quite small at $V\sin i$ = 3.5-11 km/s.

Thus, the high Li abundance in Li-rich giants is closely related to another effect, the unusually high rotation velocities of these stars.

In Table 2 the giant HR 3102 merits a separate examination. The following anomalies in the carbon and nitrogen abundances (relative to the solar abundances) have been found for it [35]: carbon deficit [C/Fe] = –0.24, nitrogen excess [N/Fe] = 0.41, and an elevated ratio [N/C] = 0.65. Anomalies of this kind for C and N are typical of giants that have passed through the CM phase (as noted in section 5.1). In this case, all the lithium must have been burnt up. The high value of log (Li) = 2.33 for HR 3120 again suggests recent synthesis of lithium.

**6.2. The super Li-rich giants.** The super Li-rich giants are a very sparse and enigmatic group. Their lithium abundances $\log \varepsilon(\text{Li}) = 3.5 - 4.3$, which substantially exceed the initial $\log \varepsilon(\text{Li}) = 3.2 \pm 0.1$, cannot be explained in terms of the standard theory of stellar evolution.

As an example, Table 3 lists the parameters of two "super Li-rich" K-giants, HR 3597+HD 77361 [53] and star No. 3416 in the Trumpler 5 cluster [54]. These stars have extremely similar $T_{eff}$, $\log g$, and $V_t$, but they have significantly different metallicity parameters [Fe/H], which for the second star is 0.5 dex below that of the sun. In addition, the mass of the second star is significantly higher than that of the first, $M$ = 1.3±0.2, at 5±1 $M_\odot$. There is no estimate of $M$ for star No. 3416 in Ref. 54; we estimated it using the same evolutionary tracks [55] employed before [53] for HR 3597. The error of ±1 $M_\odot$ in the value of $M$ for No. 3416 was obtained under the assumption that the uncertainty in log$g$ is ±0.3 dex.

TABLE 3. Parameters of Two Super Li-Rich K-Giants, HR 3597 [53] and Star No. 3416 in the Trumpler 5 Cluster

| Star | $T_{eff}$ | $\log g$ | $V_t$, km/s | $V\sin i$, km/s | [Fe/H] | $\log \varepsilon(\text{Li})$ | $^{12}C/^{13}C$ | $M/M_\odot$ |
|---|---|---|---|---|---|---|---|---|
| HR 3597 (HD 77361) | 4370 ±100 | 2.30 ±0.10 | 1.1 ±0.2 | 4.5 | -0.01 ±0.14 | 3.75 ±0.11 | 4.3±0.5[1] | 1.3 ±0.2 |
| 3416 (Trumpler 5) | 4870 | 2.05 | 1.3 | 2.8 | -0.54 ±0.10 | 3.75 | 14±3 | 5±1[2] |

[1] Kumar and Reddy [56]
[2] Our estimate for the mass M

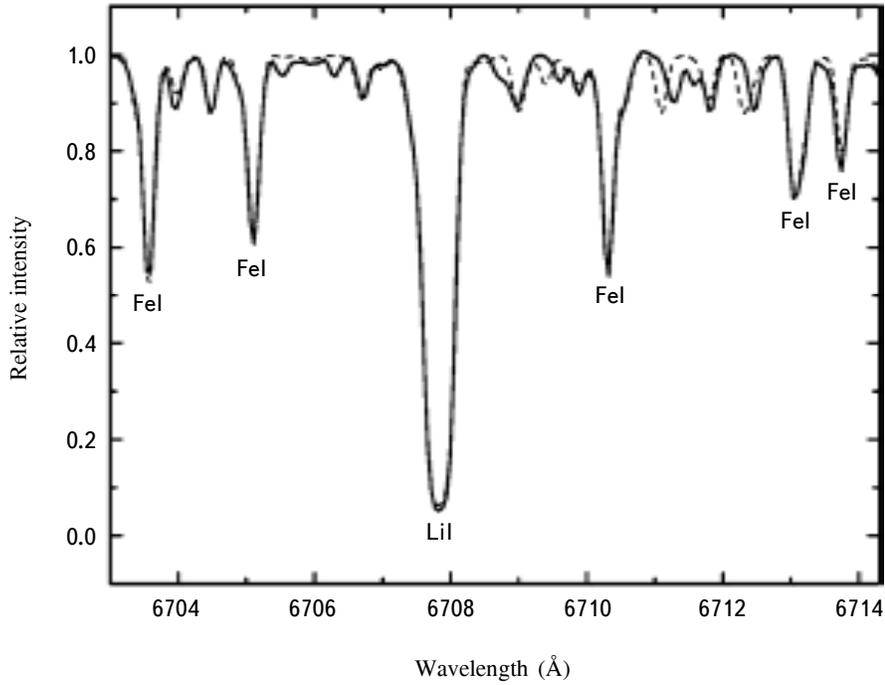

Fig. 4. A segment of a spectrum in the vicinity of the Li I 6707.8 Å line for the super Li-rich star HD 3597=HD 77361 [53]. The smooth curve shows the observed spectrum and the dashed curve, a synthetic spectrum. The agreement between the calculated and observed profiles of this strong line is reached for a high lithium abundance log (Li) = 3.69.

It is interesting that the observed rotation velocities $V\sin i$ of these two K-giants, as opposed to the Li-rich stars shown in Table 2, are extremely low and consistent with $V\sin i$ for normal K-giants.

A non-LTE analysis of the lithium abundance for the first star in Table 3 was made using three Li I lines: the 6707.8 Å resonance line and the subordinate 6103.6 Å and 8126.4 Å lines (the latter blended with molecular CN lines). For the second star, the 6707.8 Å and 6103.6 Å lines were used. Both giants yielded the same high lithium abundance, log (Li) = 3.75. Thus, the amount of Li in the atmospheres of these stars was 3.5 times the initial abundance. The high lithium abundance in the star HR 3597 is illustrated in Fig. 4 which shows a portion of the spectrum containing the Li I 6707.8 Å resonance line. Here the smooth curve represents the observed spectrum and the dashed curve, a synthetic spectrum. In order to match the calculated profile of this strong and deep line with the observed profile, a large lithium abundance of log (Li) = 3.69 was assumed [53].

These two K-giants have the important common property of a low carbon isotope ration, $^{12}C/^{13}C = 4.3\pm0.5$ for HR 3597 [56] and $14\pm3$ for No. 3416 [54]. This is very low compared to the initial (solar) ratio $^{12}C/^{13}C = 89$ [1] and is a direct proof that both of these giants have passed through the CM phase. We emphasize again that, according to the standard evolutionary theory, all of the lithium contained in the atmosphere should burn up during mixing in this phase. It has been shown [50] that a low carbon isotope ration $^{12}C/^{13}C = 4\text{-}28$ is typical of some other super Li-rich K-giants, as well as of many Li-rich giants.

**6.3. Synthesis of lithium or capture of a planet?** Modern theoretical models of stars which take rotation into account have successfully explained a number of features of cold giants and supergiants (such as the long known nitrogen-carbon anticorrelation; cf. section 5.1). They also explain the absence of lithium in the atmospheres of most stars of this type [43]. But they often fail in analyses of giants that are rich in lithium ($\log\varepsilon(Li)\geq 2$). According to the standard theory, if a star in the cold giant phase has passed through the CM phase, it should have lost all its observable lithium.

The theory can explain the high lithium abundance in the atmosphere of an FGK-giant (supergiant) only when its initial rotation velocity V0 is close to zero and, additionally, the giant has not yet reached the CM phase. When the equality $V_0 = 0$ is satisfied exactly, this kind of giant should retain its initial lithium abundance log (Li) = 3.2±0.1. If, however, $V_0$ is slightly higher, loge(Li) will be somewhat lower than its initial value. It is interesting that the observed rotation velocities for many MS B-stars, the predecessors of FGK-giants and supergiants, are actually close to zero. In particular, the peak of the distribution of the rotation velocities for MS dwarfs in classes B0-B2 lies within the interval $V\sin i = 0 - 20$ km/s (see Fig. 14 of Ref. 57). A similar result was obtained previously [58]: the maximum of the $V\sin i$ distribution for early MS B-stars lies within the range of 0-50 km/s.

In a study of the rotation velocities of MS B-stars [59] it was found that a fairly large fraction of these stars have initial velocities $V_0 \sim 50$ km/s (for details, see Ref. 43). In this regard, it should be noted that calculations [45] with a rotation model for masses $M = 2, 4,$ and $6\,M_\odot$ showed that when $V_0 = 50$ km/s, log (Li) decreases by 0.6, 1.1, and 1.4 dex, respectively, toward the end of the MS stage. This implies that stars with $M \leq 4\,M_\odot$ and $V_0 = 50$ km/s, when they reach the F-giant/supergiant stage but still have not entered the CM phase (i.e., for $T_{eff} > 5900$ K), have a lithium abundance log (Li) > 2.0, so they can be treated as Li-rich giants.

For higher initial rotation velocities (e.g., $V_0$=100 km/s), by the end of the MS stage the Li abundance in the atmosphere falls by several orders of magnitude and thereby becomes unobservable. The above discussion implies that the theoretical models can explain the existence of Li-rich stars with $M < 6\,M_\odot$ and $V_0 < 50$ km/s, but the theory is untenable for higher $V_0$.

The above scenario for the appearance of Li-rich giants may apply to some of the stars from Table 2 that are F-giants and except for one star (HR 3102), have $T_{eff} > 6200$ K, i.e., may not have reached the CM phase. This scenario does not apply to the giant HR 3102, the coldest star in Table 2 ($T_{eff} = 5690$ K) since it was found to have a deficit of C and an excess of N (see above), i.e., it has already passes through the CM phase, so that the lithium in its atmosphere should have vanished because of mixing. The standard theory also cannot explain the high rotation velocities $V\sin i$>50 km/s of some of the stars in Table 2. We recall also that many Li-rich and super Li-rich stars have a high Li abundance, but at the same time have greatly reduced carbon isotope ratios $^{12}C/^{13}C$, which prove that they have passed through the CM phase. In order to explain the high log (Li) for these stars it is necessary to assume that lithium synthesis has take place in them recently.

It is assumed that Li synthesis in red giants can take place through the Cameron-Fowler mechanism proposed [60] more than 40 years ago. It includes the reactions $^3He + \alpha \rightarrow {^7Be} + \gamma$ and $^7Be + e^- \rightarrow {^7Li} + v_e$, i.e., $^7Li$ atoms are synthesized from $^3He$ and $^7Be$. The Cameron-Fowler mechanism takes place under rather specific conditions; in particular, convection must play an important role in facilitating the rapid removal of $^7Be$ into colder layers of the atmosphere. At first this mechanism was proposed to explain the synthesis of lithium in stars with masses $M \approx 4-6 M_\odot$ in the AGB (asymptotic giant branch). Later it was used to explain the lithium enrichment of K-giants with masses $M \approx 1-2 M_\odot$ in the RGB (red giant branch), but here the Cameron-Fowler mechanism encounters certain difficulties. As opposed to the AGB stars, where ordinary convection suffices for removal of $^7Be$ to the upper layers, the giants in the RGB require extra mixing [61]. In addition, the specific segment of the RGB on which this mechanism may operate is still unclear (this problem is discussed in Refs. 50 and 63).

In recent years an alternative hypothesis has been discussed extensively: lithium enrichment of a star by capture of a giant planet such as Jupiter. Capture of this type seems entirely probable. On one hand, years of studies of exoplanets have confirmed the existence of numerous planets near red giants. On the other hand, calculations [64] show that migration of planets, which can come close to the central star, occurs in developing planetary systems. During evolution in the RGB, a red giant's radius increases, and this can disrupt the stability of planetary motion in nearby orbits. Ultimately this may cause a planet to fall into the star. This leads, first of all, to a change in the chemical composition of the star's atmosphere (including the Li abundance) and, second, it can initiate the above-mentioned extra mixing needed to trigger the Cameron-Fowler mechanism.

It has been pointed out [65] that the absorption of a large planet by a cold giant can increase the amount of lithium and also increase the star's rotation velocity. This could explain the fact, noted above, that some Li-rich giants have high rotation velocities that are utterly atypical of normal cold giants. This scenario, however, obviously cannot explain why many Li-rich giants have a low carbon isotope ratio $^{12}C/^{13}C$ which, on the other hand, is well explained by deep mixing (the CM phase) in the RGB stage.

The discovery of 20 new Li-rich giants has recently been reported by a large group of researchers [66]. A detailed analysis of their data, along with published data, made it possible to add new details to the scenario of capture of a giant planet [66]. In particular, this scenario predicts an increased frequency of Li-rich giants with increasing metallicity [Fe/H]. In fact, most such stars have normal metallicity [Fe/H]~0, while the number of stars with reduced metallicity [Fe/H]~ – 0.5 is very low.

Despite some successes in research on Li-rich and super Li-rich giants, we conclude that this phenomenon requires further study.

## 7. Lithium in magnetic Ap-stars

Magnetic Ap-stars are A- and F-stars in the Main sequence stage. Their primary feature is rather strong magnetic fields which often vary with rotational phase and reach tens of thousands of Gauss on the surface of stars (see the catalog of Ref. 67). Another feature is substantial anomalies in the abundances of many chemical elements; thus, these objects are chemically peculiar stars (CP-stars). The Li I 6707.8 Å resonance line is often observed in magnetic Ap-stars with $T_{eff}$ < 8500 K and occasionally has a variable intensity, the analysis of which yields some interesting results.

**7.1. With what should the Li abundance be compared?** The chemical anomalies of Ap-stars (relative to the sun) have an obvious trend with increasing atomic number Z: on the average, the excess is greater for heavier elements [24]. For relatively light elements, the excesses are small and may even be "negative," but with increasing Z they increase and for the heaviest elements from Pt to U (Z = 78-92), they reach 5-7 dex.

In justification of its special status in the evolution of the chemical composition of stars, it must be stated that lithium does not follow this dependence. In fact, the Li abundance is found to reach log (Li)~4 in the atmospheres of some Ap-stars (see below); that is, the excess of this light element with Z = 3 relative to the sun can be as high as ~3 dex. This large difference stands out against the background of the other light elements. This raises the question of whether comparisons with the sun are reasonable in the case of lithium.

In research on the chemical composition of stars, the sun is treated as a sort of standard star with a normal chemical composition. Over its lifetime (about 4.5 billion years), the abundances of the heavy elements in its atmosphere are essentially unchanged. In addition, they correspond to the chemical composition of the surrounding interstellar matter from which young stars continue to be born. This is proved by the agreement between the solar abundances of a number of elements with their average abundances in the atmospheres of young B-stars and AFG-giants in the vicinity of the sun [69]. Comparison with the sun as a sort of standard is fully justified from this standpoint. However, in the case of the light elements Li, Be, and B, this approach cannot be regarded as justified, since the abundances of Li, Be, and B in the sun have undergone substantial evolutionary changes. This applies especially to lithium.

As noted above in section 4, the abundance of lithium in the solar atmosphere has undergone large changes during its lifetime: from a primordial value of $\log\varepsilon(\text{Li}) = 3.2 \pm 0.1$, which is typical of young stars and the interstellar medium in the sun's vicinity, to a lower value currently estimated at $\log\varepsilon(\text{Li}) = 1.05 \pm 0.10$ [1], $1.07 \pm 0.02$ [28], and $1.03 \pm 0.03$ [70]. It is clear that the current abundance of lithium in the sun, which is a factor of 140 lower than the primordial value cannot be regarded as an initial reference point. In this case, it is better to take $\log\varepsilon(\text{Li}) = 3.2 \pm 0.1$, the initial Li abundance in young stars and in the interstellar medium near the sun, as a reference point. In the case of Ap-stars, this approach is fully justified, since we are dealing with rather young stars in the MS stage.

**7.2. Lithium spots on the surface of magnetic Ap-stars.** One characteristic feature of the spectra of magnetic Ap-stars is a regular variability in the lines of many chemical elements. This variability has been interpreted as a consequence of a nonuniform distribution of elements on the surface of a rotating star. It is analyzed using models of an inclined magnetic rotator, in other words, a rotating magnetic star whose magnetic axis (dipole axis) is inclined to its axis of rotation. Spots with elevated abundances of a number of elements have been discovered. Here the spots of some elements are concentrated near the magnetic poles, while spots of other elements may group around the magnetic equator [24].

We examine the lithium 6708 Å line in this context. The behavior of this line in the spectra of 12 magnetic Ap-stars (all of SrCrEu type) has been analyzed [71] and it was found that these stars can be divided into four groups:

*group 1* — the 6708 Å line is highly variable in intensity and in wavelength: the equivalent width of the line varies with rotation phase by more than a factor of 2 (2 stars);

*group 2* — the line is variable, but its variability is less marked (2 stars);

*group 3* — the line can be seen but is not variable (4 stars); and,

*group 4* — the line is missing in the spectrum (4 stars).

It was concluded [71] that the behavior of the 6708 Å line in groups 1-3 can be explained by the existence of two lithium spots on the star's surface in an inclined rotator model with different inclinations of the magnetic (dipole) axis to the axis of rotation. The spots lie in the regions of the magnetic poles.

For the two stars in group 1 (HD 83368 and HD 60435), it is estimated [72] that the lithium abundance in the spots is as high as $\log\varepsilon(Li) = 3.6 - 3.8$, while in the surrounding photosphere it is $\log\varepsilon(Li) = 1.8$.

Some data for the four stars in group 3, including the estimated lithium abundance [72], are listed in Table 4. All of these stars belong to the interesting roAp-type stars which undergo rapid pulsations: the pulse period $P_{puls}$ in this case ranges from 5.65 to 12.4 min (Table 4). The constancy of the 6708 Å line for these stars (its equivalent with $W \sim 100$ mÅ [71]) is explained by the following factors: first, the star's axis of rotation is directed almost toward the observer; second, the angle between the magnetic axis and the axis of rotation is fairly small. (A similar model for the star 33 Lib is illustrated in Fig. 7 of Ref. 72.) In this model of an inclined rotator, the 6708 Å line does not vary with rotational phase and the observer always sees one lithium spot located in the region of the magnetic pole. An LTE-analysis of the 6708 Å line [72] yielded an elevated lithium abundance $\log\varepsilon(Li) = 3.6 - 4.1$ (Table 4). It should be kept in mind that this value of log (Li) is averaged over the observed hemisphere of the star, while the value of log (Li) right in the lithium spot may be much higher. We note also that approximate estimates [72] indicate that these stars have an elevated lithium isotope ratio: $^6Li/^7Li = 0.2 - 0.5$ (for the sun $^6Li/^7Li = 0.08$ [1]).

The lithium 6708 Å line is missing from the spectra of the four stars in group 4. The main reason for this appears to be an excessive high effective temperature $T_{eff}$. Three of these stars are early A-stars in subclasses A2-A3 [71] and for one of them, o Vir=HD 118022, the SIMBAD data base (http://simbad.u-strasbg.fr/simbad/sim-fid) gives estimates of $T_{eff}$ = 9690-9880 K. For such high $T_{eff}$, all the lithium in the atmosphere will be ionized, so the 6708 Å line cannot be detected in principle (recall that it can be detected only for $T_{eff}$ < 8500 K). The fourth star in this group,  Cir = HD 128898, belongs to subclass A9 and has a low temperature ($T_{eff}$ = 8000 K, $\log g$ = 4.42 [73]). Calculations [43] show that if the equivalent line of the lithium line $W$ < 5 mÅ (the detection limit), then for $T_{eff}$ = 8000 K it corresponds to an upper bound for the lithium abundance of $\log\varepsilon(Li) < 2.8$; therefore, even with a rather high Li abundance, the line will not be visible. Failure to detect the 6708 Å line in the spectrum of Ap-stars can, therefore, be a consequence of an excessively high temperature $T_{eff} > 8500$ K or (for $T_{eff} < 8500$ K) of an insufficient Li abundance.

TABLE 4. Parameters of Four Magnetic roAp-Stars with a Constant Li I 6707.8 Å Line [71,72]

| HD | Name | $P_{puls}$, min | $V\sin i$, km/s | $T_{eff}$ | $\log g$ | $\log\varepsilon(Li)$ |
|---|---|---|---|---|---|---|
| 134214 |  | 5.65 | 3.0 | 7500 | 4.0 | 3.9 |
| 137949 | 33 Lib | 8.3 | 2.5 | 7750 | 4.5 | 4.1 |
| 166473 |  | 8.8 | 3.0 | 7750 | 4.0 | 3.6 |
| 201601 | γ Equ | 12.4 | 0.5 | 7750 | 4.0 | 3.8 |

As an example of a detailed study of a nonuniform chemical composition on the surface of a magnetic Ap-star (so-called Doppler mapping), let us consider charts [74] made of the distributions of 13 elements from Li ($Z = 3$) to Gd ($Z = 64$) on the surface of the star HD 3980 ($T_{eff}$=8300 K, $\log g$=4.0). It turned out that at the magnetic poles, where the field reaches 7 kG, the lithium abundance rose to $\log\varepsilon(Li) \approx 6$, or 5 dex higher than on the sun and 3 dex higher than the initial abundance in MS stars. It should be noted that, besides lithium, the concentration of such elements as O, Mn, and the rare earth elements Pr and Nd were also higher in spots near the magnetic poles. Thus, for example, the abundance of Pr in a spot was more than 5 dex higher than the solar abundance.

The above discussion shows that lithium in the atmospheres of Ap-stars is concentrated in spots at the magnetic poles, where its concentration may be several orders of magnitude higher than in the surrounding photosphere. The observed behavior of the 6708 Å lithium line (its periodic variability or constancy) depends on such geometric factors as the position of the star's axis of rotation relative to the observer or the angle of inclination of the magnetic axis to the axis of rotation.

What is the nature of the high abundances log (Li) in magnetic Ap-stars? Since other elements besides Li become highly concentrated near the magnetic poles (e.g., O, Mn, Pr, and Nd), it would be natural to search for a general cause of this phenomenon for all these elements. The most popular answer to this question is diffusion of atoms in the outer layers of the star. This hypothesis is attractive as an explanation of other features of magnetic Ap-stars, such as the above mentioned trend of an excess of elements in Ap-stars with increasing atomic number $Z$. It is also used to explain the stratification of elements with height in the atmosphere. For example, in a study [75] of the vertical distribution of a number of elements in the roAp-star 10Aql, an excess of ~4 dex was found in the abundances of Pr and Nd in high layers of the atmosphere ($\log\tau_{500} < -3.6$) relative to the deeper layers with normal abundances of Pr and Nd.

The diffusion hypothesis has been discussed briefly in Ref. 74. Here it is noted that diffusion can explain the enrichment at the magnetic poles in such elements as Li, O, Mn, Pr, and Nd, but it cannot simultaneously explain the behavior of other elements with a more complicated distribution of the star's surface. Evidently, if no alternative hypotheses appear, the diffusion theory must be greatly improved in order to fully explain the phenomena observed in magnetic Ap-stars (the trend in chemical anomalies with increasing Z, the nonuniform distribution of elements over the surface, stratification of elements with depth).

## 8. Conclusion

To sum up, we can say without exaggeration that lithium is a unique chemical element. On one hand, of all the light elements it is the most sensitive indicator of stellar evolution. On the other, it is probably the most mysterious chemical element since, as shown in this review, its observable abundance in the atmospheres of stars of different types persistently comes into conflict with theoretical predictions. Here we briefly enumerate the observational data on the abundance of lithium in stellar atmospheres and the results of comparisons with theory.

Studies of lithium in the oldest stars in the galaxy, which belong to the galactic halo, led to the discovery of the "lithium plateau." It was found that in the atmospheres of these stars the abundance of lithium is log (Li) = 2.2 and is independent of the metallicity index [Fe/H]. It is important that this value turned out to be 0.5 dex below the value log (Li) = 2.7 predicted by the standard model of the Big Bang (SMBB). Yet another conflict with theory showed up after the discovery that some stars in the halo exhibit abundances of lithium that are considerably lower than the lithium plateau, although they manifest no systematic differences from the other stars in terms of the parameters $T_{eff}$ and [Fe/H]. Neither of these conflicts have yet found a satisfactory explanation.

The very young stars in the disk in the vicinity of the sun have lithium abundances log (Li) = 3.2±0.1. This is the modern initial lithium abundance in MS stars. It is clear that this same value of log (Li) is typical of the interstellar medium, from which stars are being formed today. How did lithium enrichment of the galaxy from the primordial value log (Li) = 2.2 (the observed lithium plateau) or log (Li) = 2.7 (predicted by the SMBB) to the current level of log (Li) = 3.2 take place? Models of the chemical evolution of the galaxy show that outflows of matter from red giants, AGB stars, and novae could have contributed to this. In addition, the source of about 20% of the isotope $^7$Li in the vicinity of the sun could be spallation reactions caused by cosmic rays.

The Li abundance in the atmospheres of dwarfs in classes F, G, and K has decreased as they evolved. The drop in log (Li) is faster when the effective temperature $T_{eff}$ and mass M of a star are lower. In particular, the amount of Li in the sun's atmosphere has fallen by a factor of 140 over its lifetime (4.5 $10^9$ years). Thus far, theory has not given a convincing explanation of how the observed decrease in log (Li) took place. The reduction in log (Li) with decreasing $T_{eff}$ can be traced most clearly for stars in a single cluster (e.g., the Pleiades and Hyades clusters). The characteristic feature of the distribution of log (Li) with respect to $T_{eff}$ in old clusters is a deep dip (the lithium dip) near $T_{eff}$ ~ 6600 K. The dip may be related to internal gravitational waves, as well as to a sharp drop in the distribution of the rotation velocity $V_{rot}$ with respect to $T_{eff}$ in cold dwarfs near $T_{eff} \sim 6600$ K.

Most FGK-giants and supergiants do not exhibit the Li I 6707.8 Å in their spectra. This phenomenon is fully explained by modern models of stars that take rotation into account. The initial rotation velocity $V_0$ turns out to be an important parameter in stellar evolution; in particular, the well-known "nitrogen-carbon" anticorrelation in FGK-giants and supergiants is explained by a dependence on $V_0$. The abundance of Li depends much more strongly on $V_0$ than the abundances of N and C. In particular, for rotation velocities $V_0 > 50$ km/s, by the end of the MS stage the Li abundance in the atmosphere falls to an undetectably low level. After stars reach the FGK-giant/supergiant stage, and especially on reaching the deep convective mixing (CM) phase, a further deep drop in the atmospheric Li abundance takes place. Models of stars with masses $M < 6 M_\odot$ can explain the observed value of log (Li) only for low rotation velocities $V_0 \sim 0$; specifically, log (Li) > 1.8 up to the CM phase and log (Li)~1.5 after CM.

The FGK-giants which are rich in lithium (Li-rich giants) with an abundance $\log\varepsilon(\text{Li}) \geq 2.0$ are an especially mystifying group. All of these have masses $M < 6 M_\odot$. Theory can predict the existence of this kind of star only if it had an initial rotation velocity $V_0 \sim 0$ and, additionally, had not yet reached the deep CM phase. It is completely impossible to explain the existence of super Li-rich giants, for which the Li abundance is considerably higher than the initial log (Li) = 3.2, in terms of the standard theory. In all these cases where the theory is untenable, it is necessary to assume that lithium synthesis has recently taken place in the star. An alternative idea has been proposed: absorption of a giant planet by the star. Both of these hypotheses involve a certain amount of difficulty.

In magnetic Ap-stars with effective temperatures $T_{eff} < 8500$ K, the Li I 6707.8 Å line often varies with the rotation phase. This is interpreted in terms of the existence of lithium spots on the surface of a star. It turns out that the Li spots are found at the magnetic poles; spots of other elements, such as O, Mn, Pr, and Nd, can also be found there. In those cases where the Li I 6707.8 Å line is not variable, the explanation is related to special conditions for visibility of the spot that are determined by the orientation of the star's axis of rotation relative to the observer and by the angle of inclination of the magnetic axis to the axis of rotation. The Li abundance in a spot can be as high as $\log\varepsilon(\text{Li}) \approx 6$ or 5 dex higher than in the sun, and 3 dex higher than the initial Li abundance in MS stars. A diffusion theory is invoked to explain Li spots on the surface of magnetic Ap-stars, but it cannot provide a satisfactory explanation of all the observed features for lithium or other elements.

The author thanks N. A. Sakhibullin for a useful discussion.

**REFERENCES**

1. M. Asplund, N. Grevesse, A. J. Sauval, and P. Scott, Ann. Rev. Astron. Astrophys. **47**, 481 (2009).
2. L. S. Lyubimkov, Kinematika i fizika nebesnykh tel **26**, 32 (2010), [Kinematics and Physics of Celestial Bodies, **26**, 169 (2010)].
3. K. A. Olive and D. N. Schramm, Nature, **360**, 439 (1992).
4. K. Lind, M. Asplund, and P. S. Barklem, Astron. Astrophys. **503**, 541 (2009).
5. F. Spite and M. Spite, Astron. Astrophys. **115**, 357 (1982).

6. M. Spite, F. Spite, and P. Bonifacio, Mem. Soc. Astron. Italiana Suppl. **22**, 9 (2012).
7. A. Mucciarelli, M. Salaris, P. Bonifacio, L. Monaco, and S. Villanova, Mon. Not. Roy. Astron. Soc. **444**, 1812 (2014).
8. G. Steigman, in: C. Charbonnel, M. Tosi, F. Primas, and C. Chiappini, ed., Light Elements in the Universe (IAU Simp. 268), Cambridge Univ. Press (2010), p. 19.
9. A. Coc, S. Goriely, Y. Xu, M. Saimpert, and E. Vangioni, Astrophys. J. **744**, 158 (2012).
10. K. Lind, J. Melendez, M. Asplund, R. Collet, and Z. Magic, Astron. Astrophys. **554**, A96 (2013).
11. A. J. Korn, F. Grundahl, O. Richard, et al., Nature, **442**, 657 (2006).
12. P. Bonifacio, L. Sbordone, E. Caffau, et al., Astron. Astrophys. **542**, A87 (2012).
13. E. Caffau, P. Bonifacio, P. François, et al., Nature, **477**, 67 (2011).
14. A. Frebel, R. Collet, K. Eriksson, N. Christlieb, and W. Aoki, Astrophys. J. **684**, 588 (2008).
15. K. A. Olive, Mem. Soc. Astron. Italiana Suppl. **22**, 197 (2012).
16. S. Randich, in: C. Charbonnel, M. Tosi, F. Primas, and C. Chiappini, ed., Light Elements in the Universe (IAU Simp. 268), Cambridge Univ. Press (2010), p. 275.
17. S. C. Balachandran, S. V. Mallik, and D. L. Lambert, Mon. Not. Roy. Astron. Soc. **410**, 2526 (2011).
18. E. L. Martin, R. Rebolo, A. Magazzù, and Ya. V. Pavlenko, Astron. Astrophys. **282**, 503 (1994).
19. P. Sestito, F. Palla, and S. Randich, Astron. Astrophys. **487**, 965 (2008).
20. D. Romano, F. Matteucci, P. Ventura, and F. D'Antona, Astron. Astrophys. **374**, 646 (2001).
21. F. Matteucci, in: C. Charbonnel, M. Tosi, F. Primas, and C. Chiappini, ed., Light Elements in the Universe (IAU Simp. 268), Cambridge Univ. Press (2010), p. 453.
22. N. Prantzos, Astron. Astrophys. **542**, A67 (2012).
23. F. Matteucci, in: The Origin of the Galaxy and Local Group, Saas-Free Advanced Course **37**, 145 (2014).
24. L. S. Lyubimkov, Chemical Composition of Stars: Method and Results of Analysis, Astroprint, Odessa (1995).
25. Y. Takeda, S. Honda, T. Ohnishi, et al., Publ. Astron. Soc. Japan, **65**, 53 (2013).
26. P. Gondoin, Astron. Astrophys. **566**, A72 (2014).
27. L. S. Lyubimkov, N. S. Polosukhina, and S. I. Rostopchin, Astrophysics **34**, 65 (1991).
28. T. W. R. Monroe, J. Meléndez, I. Ramírez, et al., Astrophys. J. Lett. **774**, L32 (2013).
29. F. D'Antona, ed., The Problem of Lithium, Mem. Soc. Astron. Italiana **62**, No. 1 (1991).
30. S. G. Sousa, N. C. Santos, G. Israelian, et al., ASP Conf. **448**, 81 (2011).
31. P. François, L. Pasquini, K. Biazzo, P. Bonifacio, and R. Palsa, Astron. Astrophys. **552**, A136 (2013).
32. S. Talon and C. Charbonnel, in: C. Charbonnel, M. Tosi, F. Primas, and C. Chiappini, ed., Light Elements in the Universe (IAU Simp. 268), Cambridge Univ. Press (2010), p. 365.
33. A. Maeder, Physics, Formation and Evolution of Rotating Stars, Springer, Berlin (2009).
34. L. S. Lyubimkov, D. L. Lambert, S. A. Korotin, et al., Mon. Not. Roy. Astron. Soc. **410**, 1774 (2011).
35. L. S. Lyubimkov, D. L. Lambert, S. A. Korotin, T. M. Rachkovskaya, and D. B. Poklad, Mon. Not. Roy. Astron. Soc. **446**, 3447 (2015).

36. A. Heger and N. Langer, Astrophys. J. **544**, 1016 (2000).

37. S. Ekström, C. Georgy, P. Eggenberger, et al., Astron. Astrophys. **537**, A146 (2012).

38. L. S. Lyubimkov, Astrophysics **20**, 255 (1984).

39. Y. J. Liu, K. F. Tan, L. Wang, et al., Astrophys. J. **785**, 94 (2014).

40. O. A. Gonzalez, M. Zoccali, L. Monaco, et al., Astron. Astrophys. **508**, 289 (2009).

41. R. E. Luck, Astrophys. J. **218**, 752 (1977).

42. D. L. Lambert, J. F. Dominy, and S. Sivertsen, Astrophys. J. **235**, 114 (1980).

43. L. S. Lyubimkov, D. L. Lambert, B. M. Kaminsky, et al., Mon. Not. Roy. Astron. Soc. **427**, 11 (2012).

44. U. Frischknecht, R. Hirschi, G. Meynet, et al., Astron. Astrophys. **522**, A39 (2010).

45. U. Frischknecht, private communication (2011).

46. I. Iben, Astrophys. J. **147**, 624 (1967).

47. I. Iben, Astrophys. J. **147**, 650 (1967).

48. A. Lèbre, P. de Laverny, J. D. do Nascimento, and J. R. De Medeiros, Astron. Astrophys. **450**, 1173 (2006).

49. R. E. Luck and G. G. Wepfer, Astron. J. **110**, 2425 (1995).

50. Y. B. Kumar, B. E. Reddy, and D. L. Lambert, Astrophys. J. **730**, L12 (2011).

51. J. K. Carlberg, Astron. J. **147**, 138 (2014).

52. C. Schroeder, A. Reiners, and J. H. M. M. Schmitt, Astron. Astrophys. **493**, 1099 (2009).

53. L. S. Lyubimkov, B. M. Kaminskii, V. G. Metlov, et al., Astron. Lett. **41**, 809 (2015).

54. L. Monaco, H. M. J. Boffin, P. Bonifacio, et al., Astron. Astrophys. **564**, L6 (2014).

55. A. Claret, Astron. Astrophys. **424**, 919 (2004).

56. Y. B. Kumar and B. E. Reddy, Astrophys. J. **703**, L46 (2009).

57. S. Simón-Diaz and A. Herrero, Astron. Astrophys. **562**, A135 (2014).

58. G. A. Bragança, S. Daflon, K. Cunha, et al., Astron. J. **144**, 130 (2012).

59. H. A. Abt, H. Levato, and M. Grosso, Astrophys. J. **573**, 359 (2002).

60. A. G. W. Cameron and W. A. Fowler, Astrophys. J. **164**, 111 (1971).

61. D. Romano, F. Matteucci, P. Ventura, and F. D'Antona, Astron. Astrophys. **374**, 646 (2001).

62. F. D'Antona and P. Ventura, in: C. Charbonnel, M. Tosi, F. Primas, and C. Chiappini, ed., Light Elements in the Universe (IAU Simp. 268), Cambridge Univ. Press (2010), p. 395.

63. V. Silva Aguirre, G. R. Ruchti, S. Hekker, et al., Astrophys. J. **784**, L16 (2014).

64. W. Kley and R. P. Nelson, Ann. Rev. Astron. Astrophys. **50**, 211 (2012).

65. J. K. Carlberg, K. Cunha, V. V. Smith, and S. R. Majewski, Astron. Nachr. **334**, 120 (2013).

66. A. R. Casey, G. Ruchti, T. Masseron, et al., Mon. Not. Roy. Astron. Soc. 2016, in press (arXiv:1603. 03038v1)].

67. I. I. Romanyuk and D. O. Kudryavtsev, Astrophys. Bull. **63**, 139 (2008).

68. L. S. Lyubimkov, Izv. Krym. astrofiz. obs. **110**, 6 (2014), [Bull. Crimean Astrophys. Obs. **110**, 9 (2014)].

69. L. S. Lyubimkov, Astrophysics **56**, 472 (2013).

70. E. Caffau, H.-G. Ludwig, M. Steffen, B. Freytag, and P. Bonifacio, Solar. Phys. **268**, 255 (2011).

71. N. Polosukhina, D. Kurtz, M. Hack, et al., Astron. Astrophys. **351**, 283 (1999).

72. N. S. Polosukhina and A. V. Shavrina, Astrophysics **50**, 381 (2007).

73. P. North, S. Berthet and T. Lanz, Astron. Astrophys. **281**, 775 (1994).

74. N. Nesvacil, T. Luftinger, D. Shulyak, et al., Astron. Astrophys. **537**, A151 (2012).

75. N. Nesvacil, D. Shulyak, T. A. Ryabchikova, et al., Astron. Astrophys. **552**, A28 (2013).